%% file: cavallo.tex
\documentclass[11pt,a4paper]{article}
\usepackage{jinstpub}

\usepackage{amsmath}
\usepackage{amssymb}
\usepackage{graphicx}
\usepackage{color}

\title{Cavallo's Multiplier for in situ Generation of High Voltage}

\author[a]{S. M. Clayton,}
\emailAdd{sclayton@lanl.gov}
\author[a]{T. M. Ito,}
\author[a,1]{J. C. Ramsey,\note{Present address: Oak Ridge National Laboratory, Oak Ridge, Tennessee 37831, U.S.A.}}
\author[a,2]{W. Wei,\note{Present address: W. K. Kellogg Radiation Laboratory, California Institute of Technology, Pasadena, California 91125, U.S.A.}}
\author[b]{M. A. Blatnik,}
\author[b]{B. W. Filippone,}
\author[c]{and G. M. Seidel}

\affiliation[a]{Los Alamos National Laboratory,\\Los Alamos, New Mexico 87545, U.S.A.}
\affiliation[b]{W. K. Kellogg Radiation Laboratory, California Institute of Technology,\\
Pasadena, California 91125, U.S.A.}
\affiliation[c]{Department of Physics, Brown University,\\Providence, Rhode Island 02912, U.S.A.}

\date{March 20, 2018}

\abstract{
A classic electrostatic induction machine,
Cavallo's multiplier, is suggested for in situ production
of very high voltage
in cryogenic environments.
The device is suitable for generating a large electrostatic field
under conditions of very small load current.
Operation of the Cavallo multiplier is analyzed, with quantitative description in
terms of mutual capacitances between electrodes in the system.
A demonstration apparatus was constructed, and measured
voltages are compared to predictions based on measured capacitances
in the system.
The simplicity of the Cavallo multiplier makes it amenable to electrostatic
analysis using finite element software, and electrode shapes can be
optimized to take advantage of a high dielectric strength medium
such as liquid helium.
A design study is presented for a Cavallo multiplier in a large-scale,
cryogenic experiment to measure the neutron electric dipole moment.
}


\arxivnumber{1803.07665}

\begin{document}
\maketitle


\section{Introduction}

\input{introduction.tex}

%

\section{Cavallo's multiplier}\label{sec:cavallo_principle}

\input{cavallo_principle.tex}

\section{Demonstration apparatus}\label{sec:demo_apparatus}

\input{demo_apparatus.tex}

\section{Cavallo's multiplier in a neutron EDM experiment}\label{sec:cavallo_in_nEDM}

\input{cavallo_in_nEDM.tex}

%

\section{Discussion and Conclusion}

\input{discussion.tex}

%

\section{Acknowledgements}

This work was supported by the U.S. Department of Energy, Office of Science,
Office of Nuclear Physics contract number DE-AC52-06NA25396
under Field Work Proposal 2019LANLEEDM,
and the National Science Foundation (1506459).

\end{document}

%% file: introduction.tex

Many physics experiments require application of high voltage to
create strong electrostatic fields.
For example, electric dipole moment experiments seek
energy level shifts when a particle is subject to
a strong electric field.
High voltages are also used in some particle detectors,
such as time projection chambers (TPCs), which use an
electric field to drift charged particles toward a
detector grid.
Often a very high voltage is required in an isolated or harsh environment,
such as cryogenic, where a power supply cannot operate.

High voltage produced external to the apparatus
requires a feedthrough.
%
Cryogenic experiments, in particular those performed at sub-Kelvin temperatures,
present the additional challenge that the heat load must be small.
Feedthroughs are subject to leakage current, which can be a significant heat
input to the central part of the apparatus. In addition, the high voltage
conductors may need to be thermally anchored at heat shields. This presents an
additional challenge, as it requires thermally conducting but electrically
insulating material.
Generating the very high voltage inside the central part of the
apparatus, or multiplying a much more modest high voltage fed into the
apparatus, avoids the problems with feedthroughs.


In this paper, we suggest reviving a classic high voltage multiplier
machine, the Cavallo multiplier~\cite{cavallo1795},
to establish high voltage in cryogenic 
environments with greatly reduced demand on the voltage feedthrough.
The high voltage generator of the Pelletron~\cite{herb74},
which is based on electrostatic induction like the Cavallo multiplier and
may be thought of as its modern descendant,
is capable of higher performance and, in particular,
continuous current delivery.
However, the simplicity of the Cavallo multiplier in terms
of mechanical actuation and electrostatic design may be an
advantage in certain experiments, where only very low current is required.

The organization of this paper is as follows.  First, we explain the
principle of the Cavallo multiplier, discuss its limiting voltage gain,
solve for the voltages on the electrodes during the charging process,
and estimate heat produced by the generator.
We then show results from a room-temperature demonstration apparatus, comparing predicted
to actual voltage on the high voltage electrode throughout the charging process.
Finally, we present a design study for a Cavallo multiplier suitable for
a large-scale, cryogenic apparatus to measure the neutron electric dipole
moment.

%% file: cavallo_principle.tex
Cavallo's multiplier was invented in the late 18th century to amplify small
voltages for detection with electroscopes~\cite{cavallo1795}.
The machine uses electrostatic induction and mechanical movement to accumulate
charge onto a fixed capacitor, with a corresponding increase in voltage
across the capacitor.

\subsection{Principle of operation}

\begin{figure} 
\centering
\includegraphics[width=4.9cm]{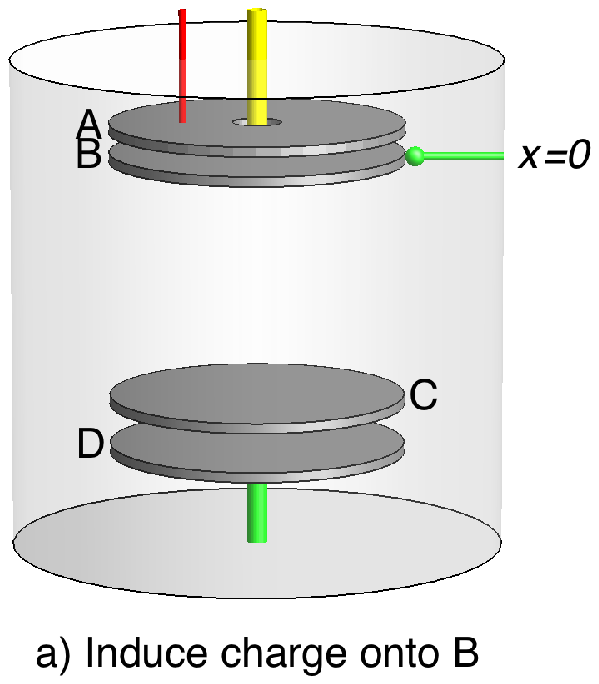}
\includegraphics[width=4.9cm]{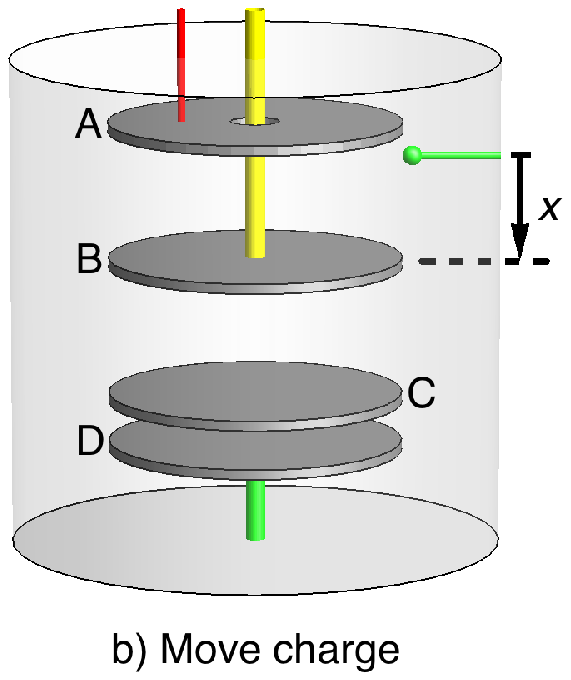}
\includegraphics[width=4.9cm]{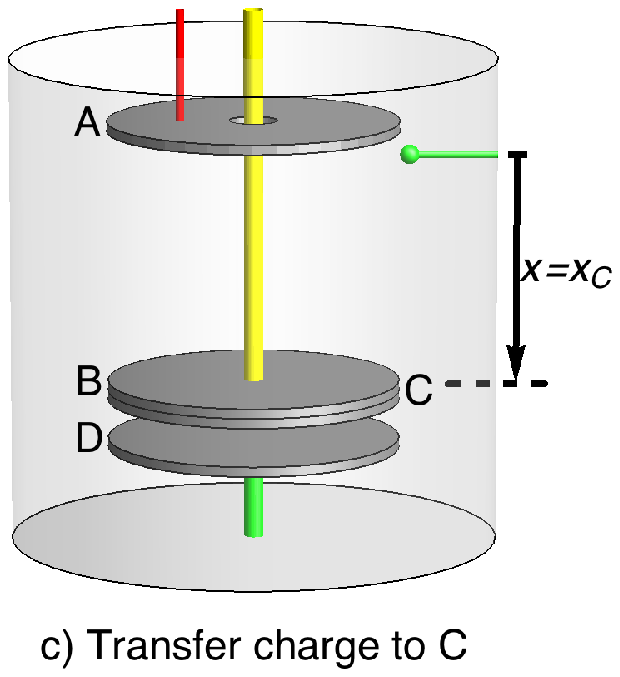}
\caption{Conceptual drawing of a Cavallo multiplier, showing the steps
 to charge the capacitor formed by the isolated electrode (C)
 and ground including (grounded) plate D.
 An external power supply is connected to plate A
 with a wire (red line in drawing) to apply a constant voltage.
 The movable plate (B) is held with an insulating rod (yellow).
 a) Plate B is grounded with a pin (top right) when positioned near plate A,
 acquiring a charge due to its capacitance with A.
 b) Plate B is disconnected from ground, which traps the induced charge,
  and is moved toward plate C.
 c) Plate B contacts plate C, transferring charge according to capacitances of the
  system.
  The steps are repeated, accumulating charge on C.
 }
\label{fig:cavallo_concept}
\end{figure}

A conceptual drawing of a Cavallo multiplier is shown in figure~\ref{fig:cavallo_concept}.
Plates A, C, and D are fixed, while plate B is movable.
Plate A is held at a modest high voltage through connection
to an external power supply.  The electrically insulated plate C
and grounded plate D comprise the capacitor to be charged to high voltage
by the apparatus.
Plate B is initially positioned near plate A, where it makes
contact with a grounding pin, acquiring an electrical charge
$Q_{B} \approx -C_{AB} V_A$ through electrostatic induction
according to its mutual capacitance $C_{AB}$ with plate A
(this expression for $Q_B$ is approximate due to
parasitic capacitance with plate C, which will be
considered below).
Plate B is moved away from A, breaking contact with the ground pin and
thus maintaining its initial charge $Q_B$.
Plate B is then brought into contact with plate C, transferring charge between the
plates.
When plates B and C are subsequently separated, the total charge $Q_B+Q_C$
will be distributed between plates B and C according to their relative mutual
capacitances with other electrodes in the system,
with the amount of charge left on plate C
\begin{equation}
Q'_C = \frac{(Q_B + Q_C) C_{CG}}{C_{CG} + C_{BG}},
\end{equation}
in which only mutual capacitances with ground G (plate D plus the ground shell)
are considered.

It may be useful to clarify the meaning of mutual capacitance
$C_{ij}$ between two conductors $i$ and $j$.
This is the constant of proportionality between the amount of
additional charge $Q_{ij}$ on conductor $i$ and its voltage
difference with plate $j$, $Q_{ij} = C_{ij} (V_i - V_j)$.
The total charge on conductor $i$ is obtained by
summing over all other conductors in the system,
$Q_i = \sum_{j\ne i} C_{ij} (V_i - V_j)$.
The total charge can also be expressed
in terms of the electric field at its surface,
$Q_i = \int_{S_i} \epsilon_0 \vec{E} \cdot d\vec{a}$
(integral form of Gauss's Law),
where the integral is over the surface of conductor $i$.
In the Cavallo multiplier,
there is no charge on the sides of plates B and C
facing each other just after they make contact because the field is zero
between the plates;
we see that plate C ``shields'' B from plate D, leaving only
the much smaller capacitance to the ground shell
to contribute to $C_{BG}$.

Another cycle begins by moving B back to its
position near A and in contact with the grounding pin.
Starting with $Q_C = 0$, after $n$ cycles
the charge accumulated onto C is
\begin{equation}
Q_{Cn} = Q_B \frac{\epsilon (1 - \epsilon^n)}{1 - \epsilon},
\end{equation}
where $\epsilon \equiv C_{CG}/(C_{CG} + C_{BG})$.
The amount of charge loaded onto plate C
in the limit of infinite number of cycles is
\begin{equation}
\lim_{n \to \infty} Q_{Cn} = Q_B \frac{C_{CG}}{C_{BG}},
\end{equation}
equivalent to the condition of no net charge transferred to
C after contact with B ($Q'_B = Q_B$).
This amounts to a maximum voltage achievable on plate C of
$V_C^{\rm max} = -V_A C_{AB}/C_{BG}$, independent of the capacitance $C_{CG}$.

As the total amount of charge on plate C increases, it may be important
to account for its influence on the initial charge loaded onto plate B,
\begin{equation}\label{eq:QB0}
Q_B^a = -C_{AB}^a V_A - C_{BC}^a V_C^a,
\end{equation}
where the superscript $a$ indicates
that parameters correspond to when B is in its initial position next
to A.
Similarly, we can account for the influence of plate A on the
charge that remains stuck on B, when B is in contact with C,
\begin{equation}\label{eq:QB1}
Q_B^c = C_{AB}^c (V_C^c - V_A) + C_{BG}^c V_C^c,
\end{equation}
where the superscript $c$ labels this position.
The condition for maximum possible charge on C is when all of the
charge loaded onto B remains stuck on the B electrode,
$Q_B^a = Q_B^c$.
Equivalently, the charge on plate C is unchanged, $Q_C^c = Q_C^a$,
giving an additional expression,
\begin{equation}\label{eq:QC01}
Q_C = C_{CG}^c V_C^c = (C_{CG}^a + C_{BC}^a) V_C^a.
\end{equation}
Combining eqs.~\ref{eq:QB0}--\ref{eq:QC01} and $Q_B^a = Q_B^c$
gives an expression for the maximum voltage on
C (when B remains in contact with C) relative to the voltage $V_A$ fed
into the system,
\begin{equation}\label{eq:max_Vc}
-\frac{V_C^{\rm1,max}}{V_A} = \frac{C_{AB}^a - C_{AB}^c}{C_{BG}^c + C_{AB}^c
 + \kappa C_{BC}^a } \equiv G^{\rm max},
\end{equation}
where $\kappa \equiv C_{CG}^c/(C_{CG}^a+C_{BC}^a)$
and we define the maximum voltage gain $G^{\rm max}$.
Typically, $\kappa \approx 1$ and $G^{\rm max}$ does not depend
on the load capacitance $C_{CG}$.

To discharge plate C, $V_A$ is reduced, zeroed, or reversed depending
on how much charge is to be removed in a given cycle.
If $V_A = 0$, $V_C$ approaches zero asymptotically with number
of cycles.  Fully discharging plate C requires setting $V_A$
to its slightly reversed value (same polarity as plate C) to
remove the last bit of charge.

%
The machine in figure~\ref{fig:cavallo_concept} can be further understood
by generalizing eqs.~\ref{eq:QB1} and~\ref{eq:QC01} to any plate B position $x$,
\begin{eqnarray}
Q_B &=& C_{AB} (V_B - V_A) + C_{BC} (V_B - V_C) + C_{BG} V_B,\label{eq:Q_B} \\
Q_C &=& C_{BC} (V_C - V_B) + C_{CG} V_C,\label{eq:Q_C}
\end{eqnarray}
in which the mutual capacitances and voltages are functions of $x$.
We neglect the mutual capacitance between the fixed plates A and C.
%
%
%
%
%
%
The solutions for $V_B$ and $V_C$ with given $Q_B$ and $Q_C$ are
\begin{eqnarray}
%
%
V_B &=& \frac{ Q_B + C_{AB} V_A + \eta Q_C }
  { C_{AB} + C_{BG} + \eta C_{CG} },\label{eq:V_Bx} \\
V_C &=& \frac{ Q_C }{C_{CG} + C_{BC}}
                               + \eta V_B\label{eq:V_Cx},
\end{eqnarray}
with $\eta \equiv C_{BC}/( C_{CG} + C_{BC} )$.

As plate B approaches plate C to within an infinitesimal gap, we expect
their mutual capacitance $C_{BC}$ to diverge to infinity.
In this limit, as evident
from eq.~\ref{eq:V_Cx}, plates B and C approach the same voltage prior to touching.
The electric field between the plates during the approach depends on
details of the geometry.  If the plates are parallel planes of infinite
extent, the field between them is
\begin{eqnarray}
E_{BC}(x) &=&
      \frac{V_B(x) - V_C(x)}{\Delta x} \\
 &=& ( (1 - \eta(x)) V_B(x) - \frac{ \eta(x) Q_C }{ C_{BC}(x) })/\Delta x \\
 &=& \frac{V_B(x) C_{CG} - Q_C}{(C_{CG} + C_{BC}(x)) \Delta x},\label{eq:E_BC}
\end{eqnarray}
where $\Delta x = x_C - x$ is the distance between plates B and C.
To calculate the field just before the plates touch, we take the limit
of eq.~\ref{eq:E_BC} as $\Delta x \to 0$ and $C_{BC} \gg C_{CG}$,
%
%
\begin{equation}\label{eq:E_BC_limit}
\lim_{x \to x_C} E_{BC}(x) =
\lim_{x \to x_C}
\frac{V_B(x) C_{CG} - Q_C}{C_{BC}(x) \Delta x}.
\end{equation}
Applying the same limit to eq.~\ref{eq:V_Bx},
we have
\begin{equation}
\lim_{x \to x_C} V_B(x) = \frac{Q_B + Q_C + C_{AB} V_A}{C_{AB} + C_{BG} + C_{CG}},
\end{equation}
and the numerator in eq.~\ref{eq:E_BC_limit} is finite as the plates approach.
The electric field is finite just before contact
if $C_{BC}$ diverges as $(\Delta x)^{-1}$ (or faster),
a condition met by a parallel plate capacitor:
$C_{BC}(\Delta x) = \epsilon A/\Delta x$,
with plate area $A$ and dielectric constant between the plates $\epsilon$,
giving
\begin{equation}\label{eq:Emax_PP}
\lim_{x \to x_C} E_{BC}(x) = \frac{V_B(x_C) C_{CG} - Q_C}{\epsilon A}.
\end{equation}
The actual field before contact between real electrodes depends on details of
the surfaces, electrode shapes and their alignment.
In general, we expect the field to diverge prior to contact and create a spark.

\subsection{Heat Generation}

The electrostatic potential energy of the charged system is provided
by the mechanical energy required to move plate B between plates A and C,
plus the external power supply sourcing or sinking charge on plate A.
Here, we consider heat generation in the Cavallo multiplier itself,
not including work done by the external power supply.
There are three possible sources of dissipation in this otherwise
efficient process: sparking between electrodes, heat from charges
flowing across conductor surfaces, and drag from moving plate B
through a fluid medium (unless in vacuum).

Due to the high electric field as plate B approaches C,
at some point prior to contact charge will flow between the plates
via spark or field emission and dissipate energy in the process.
The maximum energy available to the spark is the difference in electrostatic
configuration energy before and after the event,
$W_{\rm spark} = W_{\rm initial} - W_{\rm final}$,
with the configuration energy given by
\begin{equation}\label{eq:config_energy}
W = \frac{1}{2} \sum_i Q_i V_i,
\end{equation}
and the sum is over all electrodes in the system.
This energy is the upper limit on heating due to sparks between
the electrodes and will be estimated with a parallel plate model
in Sec.~\ref{sec:parallel_plate}.
There can also be a spark as plate B returns to $x = 0$ and contacts
the grounding pin, though this can be suppressed by temporarily reducing
$V_A$ to an appropriate setting depending on the residual charge $Q_B^c$
left on B, before B contacts the ground pin.

Another source of heat is due to rearrangement of charges on electrodes
with non-zero surface resistivity $\rho_S$, for example from the top to bottom surfaces
of plate B as it moves from initial proximity to plate A down to plate C:
\begin{eqnarray}\label{eq:heat_from_currents}
W_{\rm q} &\sim& \Delta t I^2 \rho_S \nonumber \\
          &\sim& \Delta t (Q_B/\Delta t)^2 \rho_S \nonumber \\
          &\sim& \rho_S Q_B^2/\Delta t,
\end{eqnarray}
where $\Delta t$ is a characteristic time interval over which charge flows
from the top to bottom of the plate.
Similarly, as plate B approaches C, charges are repelled from the side of C facing B,
leaving the surface facing B with charge approaching $-Q_B$ in the limit of B
and C almost touching.  The amount of heat generated can be estimated from
eq.~\ref{eq:heat_from_currents} with appropriate characteristic $\rho_S$ and $\Delta t$.

%
If the multiplier is immersed in a classical fluid, heat is generated by the drag
force on the movable plate B. If the flow about the plate is laminar, the drag
force depends on the shape and dimensions of the plate and properties of the
fluid, and it is proportional to the velocity. Hence the rate of energy dissipation
depends on $v^2$. For the more likely case of turbulent flow the power loss  is
dependent on $v^3$,
\begin{equation}\label{eq:drag}
P_D = F_D v = c_D A \rho v^3/2,
\end{equation}
where $A$ is the plate area,  $\rho$ is the fluid density and $c_D$ is the drag
coefficient that depends upon fluid properties and object shape. The case of a
Cavallo multiplier operating in superfluid helium is discussed below
in Sec.~\ref{sec:cavallo_in_nEDM}.

\subsection{Parallel-Plate Capacitor Approximation}\label{sec:parallel_plate}

\begin{figure} 
\centering
\includegraphics[width=10cm]{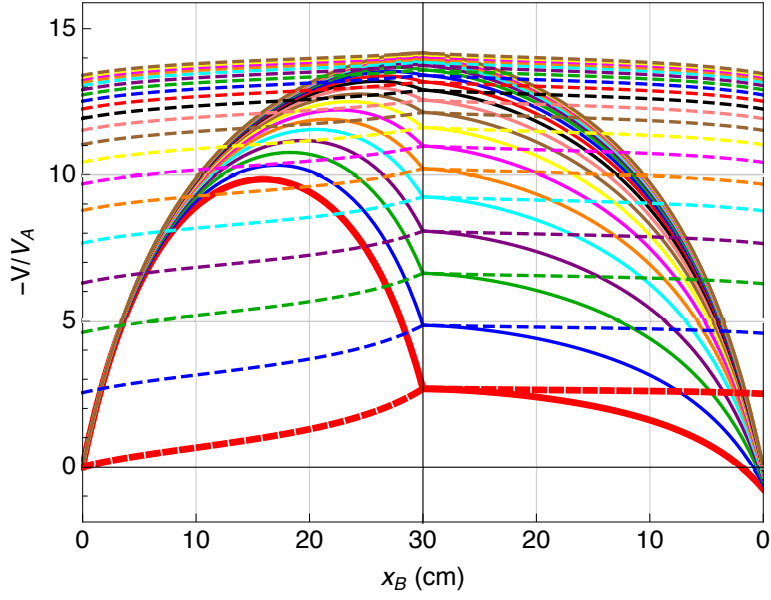}
\caption{
 Voltages on plates B (solid lines) and C (dashed lines) in the simple model
 using the ideal parallel plate capacitance
 formula, over several charging cycles, starting with zero initial
 charge on plate C.  A cycle starts on the left ($x_B=0$) with
 plate B grounded, proceeds to the right with B-C contact at $x_B = 30$~cm,
 and continues to the right as plate B moves away from C after contact,
 returning to $x_B=0$ with a non-zero voltage prior to grounding.
 The thicker lines correspond to the first cycle.
}
\label{fig:charging_voltages}
\end{figure}

\begin{figure} 
\centering
\includegraphics[width=10cm]{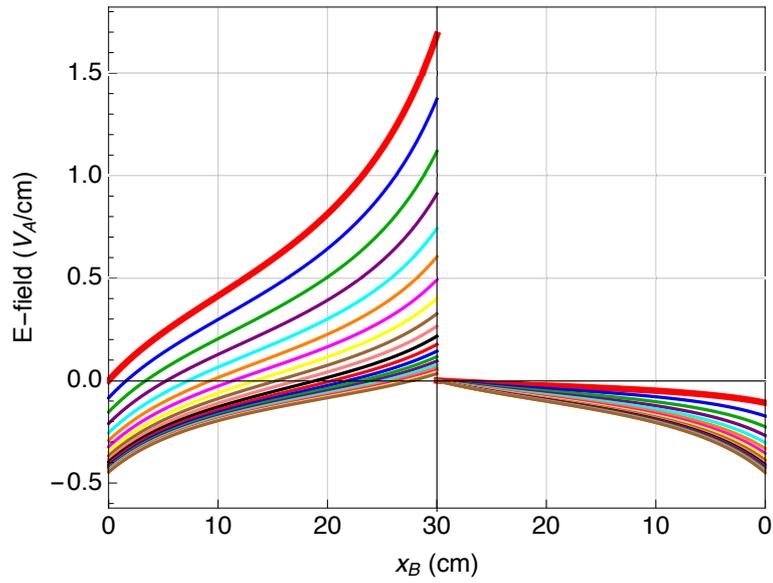}
\caption{
 Electric field between plates B and C in the
 same model as in figure~\ref{fig:charging_voltages},
 with the same horizontal axis.
 The highest field occurs in the first cycle
 (plotted as a thicker line), and falls off in
 subsequent cycles.
}
\label{fig:charging_fields}
\end{figure}

\begin{figure} 
\centering
\includegraphics{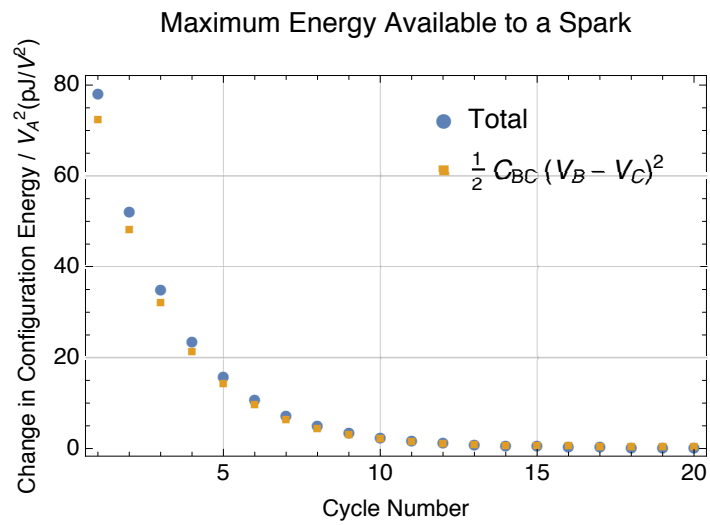}
\caption{
 Energy available to a spark in the simple parallel-plate capacitor model,
 based on change in the configuration energy (eq.~\ref{eq:config_energy})
 at 0.5~cm separation between plates B and C, if short-circuit suddenly
 develops between these plates.
 The blue points are the total change in energy, and the orange points
 are $C_{BC} (V_B - V_C)^2/2$ evaluated at the same 0.5~cm separation.
}
\label{fig:spark_energy}
\end{figure}

For further insight into the physics of Cavallo's machine, we model the mutual
capacitances between the plates with the ideal parallel plate capacitor
formula,
\begin{equation}
C_{ij} = \frac{\epsilon_0 A}{|x_i - x_j|}.
\end{equation}
Results of applying the above analysis to this model are shown in
figures~\ref{fig:charging_voltages} through~\ref{fig:spark_energy} for the
following parameters:
diameter of plates $d = 40$~cm, initial B-C distance (stroke length
of the moving plate B) $L = 30$~cm,
capacitance of the target electrode to ground $C_{\rm CG} = 70$~pF,
parasitic capacitance of plate B to ground $C_{\rm BG} = 8$~pF (constant
value; in a real system this would likely be lower when B is near C).

 Plate B obtains a high voltage on its way to C
 as its capacitance with other electrodes decreases (figure~\ref{fig:charging_voltages}).
 In a real system, the ground shell needs to be
 sufficiently far away from B to prevent electrostatic
 discharge due to too strong of an electric field.
 As plate B nears C, the voltages approach the same value
 before the plates contact.
 In this idealized model, the electric field between B and
 C steadily increases toward a finite value (figure~\ref{fig:charging_fields}).
 In practice, a surface feature or edge effect would create a localized 
 high field and eventually a spark.
 A spark is modeled as a short circuit between B and C ($V_B = V_C$,
 and $Q_B + Q_C$ conserved).
 The maximum energy available to a spark,
 the change in total electrostatic potential energy (eq.~\ref{eq:config_energy}),
 is plotted in figure~\ref{fig:spark_energy} assuming the spark occurs at
 5-mm B-C gap in each cycle.
 At this gap, the B-C capacitor accounts for most of the energy available to
 a spark, with the rest attributable to parasitic capacitances.

 The highest field between B and C is on the first cycle,
 when C is initially uncharged.  The highest spark energy is
 correspondingly on the first cycle.  The charging voltage $V_A$
 could be reduced for the first few cycles, allowing B and C to
 approach closer before the electric field exceeds some breakdown
 condition.  The net energy deposited into sparks could thus be
 reduced.  There is also a spark upon return of plate B to its
 position next to A, when it makes contact with the ground pin.
 In this case, $V_A$ could be adjusted (reduced) before B contacts
 ground to suppress the spark, and subsequently reset to its nominal
 value to more gently induce charge onto B.

%% file: demo_apparatus.tex
We constructed an apparatus to test quantitative understanding of
the voltage development on plate C.  The apparatus geometry is
similar to that shown in figure~\ref{fig:cavallo_concept}, with electrode
diameters of approximately 40~cm and stroke length 29~cm.
The Plate C electrode is supported by G10 insulators on a standoff
ring, which was connected to common ground in these studies.
The moving electrode, plate B, is supported by a solid
G10 rod which passes through the top shelf to a linear actuator.
Before moving away from its upper position next to plate A,
plate B is electrically isolated from ground by retracting a
solenoid-actuated grounding pin.
\begin{figure} 
\centering
\includegraphics[width=0.45\textwidth]{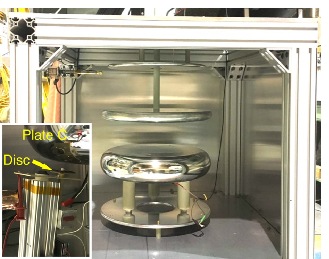}
\hspace{5mm}
\includegraphics[width=0.35\textwidth]{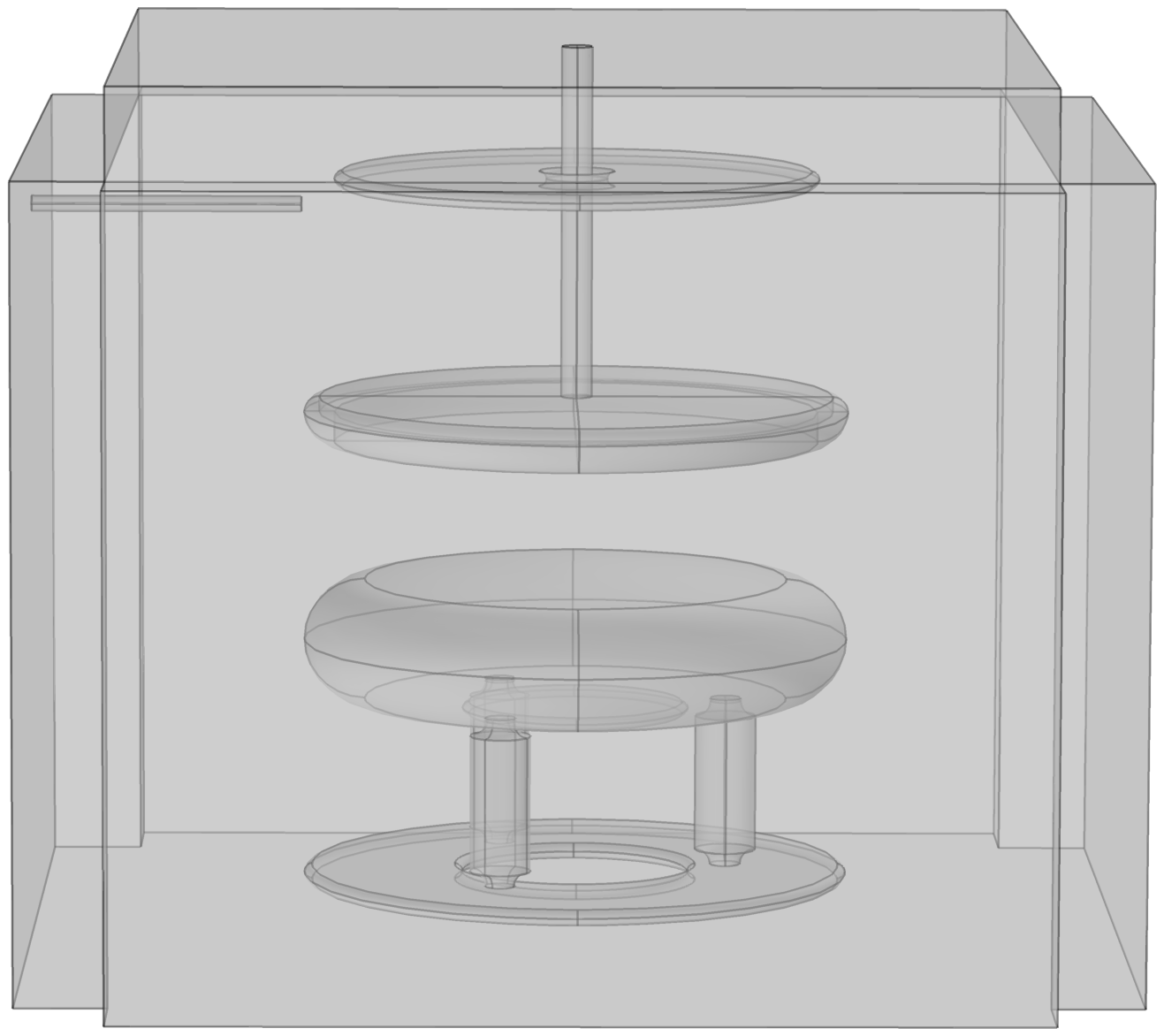}
\caption{
Left: photograph of the demonstration apparatus, with two of the
grounded sides removed.  Here, plate B is about halfway along its path to
plate C.  The vibrating disc, non-contact electric field probe,
not present in the main photograph but shown in the inset photograph,
was positioned below and at the outer edge of plate C.
Right: COMSOL~\cite{comsol} geometry used to compute mutual
capacitances between the electrodes in the demonstration apparatus.
}
\label{fig:demo_photo}
\end{figure}
A photograph of the apparatus is shown in figure~\ref{fig:demo_photo},
along with the apparatus geometry representation in the finite element
software COMSOL~\cite{comsol},
which was used to check measured capacitance values.
Not present in the photograph and COMSOL model is an additional
grounded electrode immediately underneath Plate C to increase load capacitance.

\begin{figure} 
\centering
\includegraphics[width=8cm]{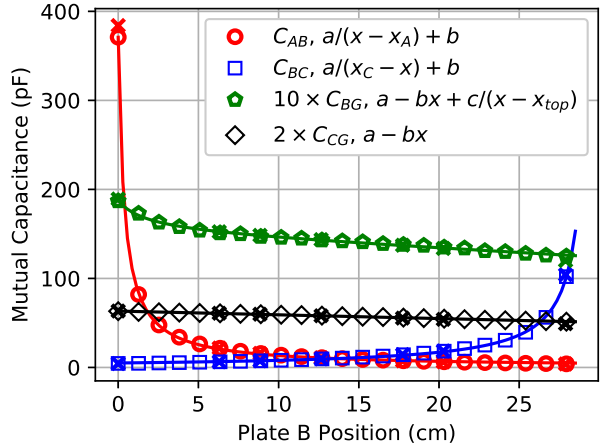}
\caption{
Mutual capacitances between electrodes in the demonstration apparatus,
measured and calculated with COMSOL~\cite{comsol} (exes).
Fits to the measured values by functions listed in the legend are shown,
where $x$ is the horizontal axis and all other parameters are fitted.
The data and fits for $C_{BG}$ and $C_{CG}$ are scaled for clarity in the plot.
An additional grounded plate underneath Plate C was added later,
which increased $C_{CG}$ by a constant $\approx$100~pF.
}\label{fig:demo_capacitances}
\end{figure}
%
The mutual capacitances between pairs of electrodes in the system
(including ground) were measured as a
function of plate B position using a handheld capacitance meter
(Agilent U1733C).
A pair of coaxial cables served as measurement leads, with
the outer conductors connected to the guard input on the meter.
These cables were passed through the top deck of the apparatus,
and electrodes $i$ and $j$ were connected to the meter inputs
via the coaxial cables center conductors to measure $C_{ij}$.
All electrodes in the system other than $i$ and $j$ 
were also connected to the guard input of the meter,
and all grounded sides of the apparatus were in place during measurements.
This arrangement was effective in removing parasitic capacitances,
ensuring the measured value was the mutual capacitance $C_{ij}$.
Results of these measurements are shown in figure~\ref{fig:demo_capacitances}
along with fits to empirical functions.  The $C_{ij}$ calculated with
finite element software are also plotted, showing good agreement
with the measurements.

At the relatively small plate charges in the demo apparatus for reasonable
applied voltages on Plate A, leakage current must be minimized.
We employed a non-contact method to measure the voltage by measuring the current
amplitude from a small ($\approx$8~cm$^2$) vibrating disc in the fringe field of
Plate C.
The disc was positioned below Plate C near the outer edge
(see figure~\ref{fig:demo_photo} inset), where
the influence of field lines from other non-grounded conductors is expected
to be negligible.
The current from the disc was amplified by a high-gain current-to-voltage
preamplifier (Stanford Research SR556), which maintains the disc at
virtual ground potential, followed by a lock-in amplifier (Stanford Research SR830).
The reference oscillator from the lock-in amplifier was input into
a high voltage operational amplifier (TI OPA548) driving
a mechanically-amplified piezo-actuator (Thorlabs PK2FVF1), with parameters set to
move the disc with $\approx$40~$\mu$m amplitude, 40~Hz sinusoidal displacement.
In this arrangement, the lock-in out-of-phase component
was proportional to the voltage on Plate C.
The constant of proportionality between the lock-in signal and $V_C$ was
calibrated by connecting an electrometer to Plate C and running a Cavallo
charging cycle to sweep out $V_C$ voltages.

We note that a possible alternative non-contact electric field transducer
for this measurement is the ``field mill,'' which replaces the vibrating
capacitor with a fixed capacitor that is alternately guarded from or exposed
to the electric field by a rotating vane.
That type of arrangement likely offers a much larger raw signal,
but the vibrating capacitor method may be 1) more readily adapted to the
strictly non-magnetic, sub-1 Kelvin cryogenic environment of the
experiment discussed in Sec.~\ref{sec:cavallo_in_nEDM};
and 2) with suitable electrostatic design, less likely to cause sparking
when operating in a high field region.
However, with respect to point 1,
a field mill variant that may be suitable for cryogenic environments
uses a micromachined vibrating aperature~\cite{horenstein2001}.

\begin{figure} 
\centering
\includegraphics[width=8cm]{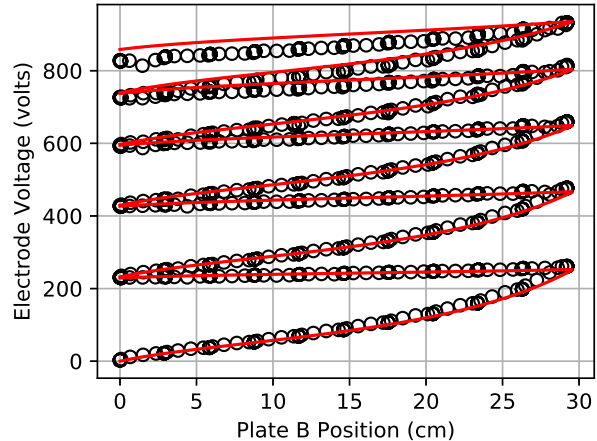}
\caption{
Voltage on the demonstration apparatus plate C during five
charging cycles with $V_A = -100$~volts.
Plotted are the measured voltages with the non-contact
vibrating capacitor method (circles) and
predicted voltages (line) based on
the fits to measured mutual capacitances (figure~\ref{fig:demo_capacitances}).
}\label{fig:demo_voltageData}
\end{figure}

With the non-contact voltage measurement calibrated, the electrometer was
disconnected, and several charging cycles were performed.  Results are shown in
figure~\ref{fig:demo_voltageData}, along with predicted voltages based on
the fits to measured mutual capacitances (figure~\ref{fig:demo_capacitances})
and application of eqs.~\ref{eq:V_Bx} and~\ref{eq:V_Cx}.
Voltages in the first few cycles agree well with prediction, with some systematic
disagreement especially in the last cycle.
Possible causes of the disagreement include
the cumulative effect of errors in capacitance measurements
and leakage current through insulating standoffs.
Although care was taken in the apparatus construction
to avoid the triboelectric effect caused by the rod sliding against
dissimilar material (such as a plastic bearing block in an earlier iteration),
another possible source of systematic error is from charges stuck to the
G10 rod holding Plate B, which would impact the charge initially
induced on Plate B and that transferred to Plate C.
At much higher operating voltages, this effect should be relatively
less important.  Leakage current through the insulators
supporting plate C could in principle be accounted for by connecting the
standoff ring to the input of a picoammeter (virtual ground), rather
than directly to ground, and logging these data to enable a correction
in the data analysis; however, this was not done in the present study.
In any case, it appears the voltages in the demonstration apparatus
can be reasonably well predicted from measured or calculated capacitances,
validating an approach of using calculated capacitances
to evaluate designs for a production apparatus.

%% file: cavallo_in_nEDM.tex
As a specific example of incorporating a Cavallo multiplier into a
physics experiment, we performed a high voltage design study for
a large-scale, cryogenic experiment to search for the
neutron electric dipole moment (nEDM)~\cite{gollam,ito_nedm}.
The central detector is a large ($\sim$1~m$^3$) volume containing
liquid helium at about 0.4~Kelvin.
A capacitor is formed by a central high voltage electrode flanked
on either side by measurement cells followed by ground electrodes.
To take advantage of the excellent dielectric strength of liquid
helium, a design goal is to charge this capacitor,
with $C_{CG} \approx 100$~pF, to about 650~kV, corresponding to
a field inside the measurement cells of 75~kV/cm.


Bringing 650~kV into a volume filled with LHe at 0.4~K with a direct
HV feed is a very challenging task. The HV feed line needs to be
designed so that the heat brought in is within the allowed heat budget
of a few tens of milliwatts, which requires proper thermal anchoring at 4~K and
77~K. In addition, the feedthrough needs to be superfluid tight
and, for the nEDM experiment, strictly non-magnetic.
The leakage current flowing across the surface of the insulator must
be minimized to avoid heating. Note that a leakage current of 15~nA flowing across
the surface of a 650~kV HV feedthrough insulator would cause a heating of
10~mW.
Bringing in 50~kV is relatively easy, and compact (few centimeter scale)
commercial feedthroughs are available with
$\sim$1~nA leakage current~\cite{mshv,nonmag_ft};
therefore, we set $V_A = 50$~kV.  A stable field of 100~kV/cm in liquid
helium with large electrodes has been demonstrated~\cite{mshv},
so we set this as the maximum allowed field in this design study.

\begin{figure}
\centering
\includegraphics[width=8cm]{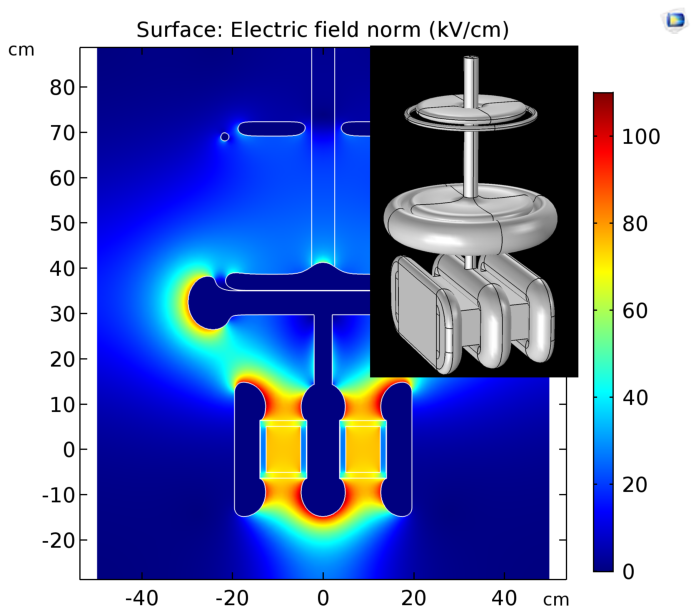}
\includegraphics[width=6cm]{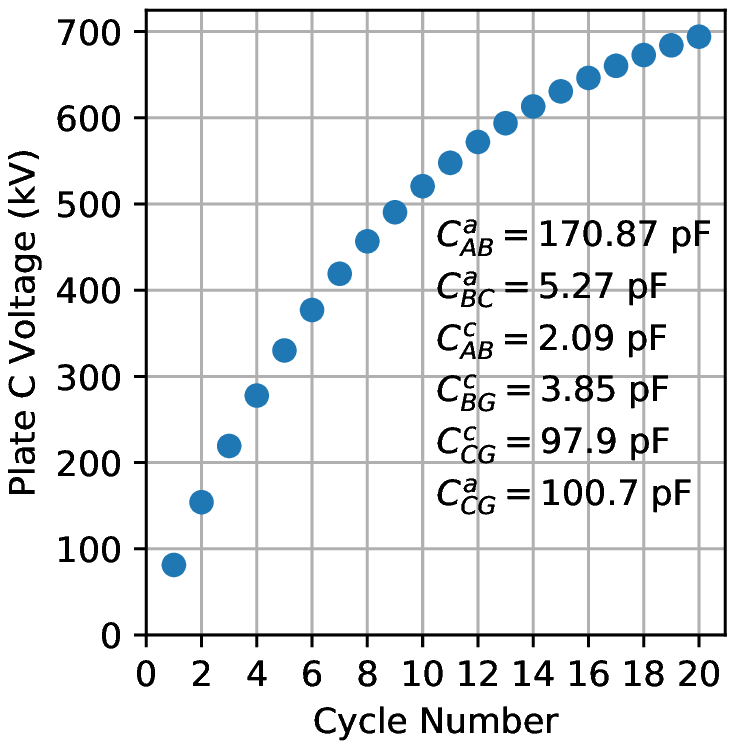}
\caption{
 Full 3D COMSOL~\cite{comsol} simulation of the nEDM
 central volume including a ground return at 50~cm radius.
 Left: electric field strength at the midplane when C is fully charged to
 650~kV and plate B remains in contact with C
 (the large electrode in the middle).
 The highest field around plate C is about 80~kV/cm.
 Right: calculated $V_C$ vs. number of charge cycles completed,
 using the computed mutual capacitances and application of
 eqs.~\ref{eq:V_Bx} and~\ref{eq:V_Cx}.
}
\label{fig:comsol_nEDM}
\end{figure}
%

The starting point is a Cavallo multiplier with geometry similar
to figure~\ref{fig:cavallo_concept},
with a conducting rod between plate C and the central high voltage electrode in
the measurement cell structure.
The maximum achievable voltage gain $G^{\rm max}$ (eq.~\ref{eq:max_Vc}),
obtained asymptotically with increasing number of charging cycles,
is a useful guide for choosing an initial electrode geometry.
In a practical application with a reasonable number of cycles,
we could charge the system until for example $\approx$90\% of the maximum gain
is reached; thus, our goal is $G^{\rm max} = 13/0.9 = 14.4$.
Consistent with a maximum electric field of 100~kV/cm, the initial
A-B gap is set to 5~mm, with appropriate
Rogowski profiles~\cite{rogowski,cobine}
for plates A and B to avoid high electric field at the plate edges.
Increasing $C_{AB}^a$, without increasing the electric field strength,
requires increasing the plate area.
The other capacitances in eq.~\ref{eq:max_Vc} should be minimized
for the highest voltage gain.
Increasing the plate B stroke length ($x_C$) decreases
$C_{AB}^c$ and $C_{BC}^a$.
The remaining significant capacitance affecting the gain
is $C_{BG}^c$, the parasitic capacitance of B to ground when next to
plate C.  This can be reduced by adding lobes to plate C to partially
envelop B, partially shielding it from ground.

We performed a finite-element study using COMSOL~\cite{comsol} to find a nominal
geometry appropriate for the nEDM apparatus.
The purpose was to evaluate mutual capacitances between conducting bodies
and look for electric field hot spots, and several iterations were performed.
Given the capacitances from COMSOL evaluated when B is at its initial (next to A)
and final (touching C) positions, the formulae from the analysis
in section~\ref{sec:cavallo_principle} can be applied.
This is shown in figure~\ref{fig:comsol_nEDM}, the voltage on the C electrode
versus charging cycle number.
The electric field distribution when C is charged to 650~kV is shown
in figure~\ref{fig:comsol_nEDM}.  The A and B electrodes have diameter 40~cm,
and the part of the C electrode in the middle of the figure~\ref{fig:comsol_nEDM}
has lobes around its perimeter to reduce $C_{BG}^c$.
%

For this design to be feasible, the heat load must be small,
less than about 10~mW averaged over a $\sim$100-second total charging period.
The heat energy input due to charge flow across resistive electrodes 
appears to be negligible: $W_q \sim \rho_S Q_B^2 / \Delta t \sim 1$~nJ
per cycle, with plate resistivity $\rho_S \sim 1$~k$\Omega/\square$,
initial charge $Q_B \sim 1$~$\mu$C, and characteristic charge
movement time $\Delta t \sim 1$~s.
For the energy deposited by sparks, the parallel plate toy model
in section~\ref{sec:parallel_plate} is relevant, which
shows a spark energy (figure~\ref{fig:spark_energy})
of at most $W_s \sim (80\ {\rm pJ/V^2})\times (50\ {\rm kV})^2 = 0.2$~J
in the first cycle and falling off quickly in subsequent cycles.
If needed, $W_s$ can be reduced by lowering $V_A$ for the first few
cycles, when the field between plates B and C would otherwise be
much higher than in later cycles~(see figure~\ref{fig:charging_fields}),
possibly allowing B to approach closer to C before sparking.
Constructing the electrodes with designated contact areas made
of high dielectric strength, bulk-resistive material could decrease
the instantaneous power of the sparks by slowing charge transfer
between electrodes, possibly avoiding electrode surface damage
but not reducing the total heat production due to charge transfer.

%
The motion of the B electrode in superfluid helium at 0.4 K can
create heat by two mechanisms. Firstly, a drag force arises from the motion of an
object through the thermal phonon excitations existing in the liquid, the classical
analogue of dissipation in laminar flow. And secondly, when the velocity of the
object reaches a critical value, quantum turbulence, an entanglement of quantized
vortices which has no classical correspondence, is generated. Measurements have
been made of these phenomena at low temperature by the motion of  oscillating
spheres, wires, grids and tuning forks~\cite{Vinen} but at sizes of millimeters
or smaller. Results obtained at these very small scales must be extrapolated by
orders of magnitude to obtain estimates for heat produced by the motion of the
40~cm diameter movable electrode.

The experiments by Niemetz and Schoepe~\cite{niemetz04} on a 124~$\mu$m oscillating
sphere are perhaps the most directly related to the case of a moving plate. From
their data for the phonon drag force at 0.3~K, which has a magnitude of piconewton-scale,
the estimated heat produced by the 40-cm diameter electrode with a speed of 3~cm/s in
helium at 0.4~K is the order of 20~$\mu$W and is no concern. However, the
velocity of 3~cm is well above the critical velocity for the generation of quantum
turbulence by the electrode~\cite{Kopnin}. The drag force, $F_D$, from quantum
turbulence is found to vary with velocity, $v$, approximately as $F_D\approx \gamma v^2$,
where the coefficient, $\gamma$, when expressed in the form of the drag coefficient
for classical turbulence, eq.~\ref{eq:drag}, is $\gamma = c_D \rho A/2$. The fact that the
drag from quantum turbulence can be characterized by the same general expressions
as for classical hydrodynamics provides some confidence that the estimated heat
production by the electrode, obtained by extrapolating measurements made on
objects 3 orders of magnitude smaller in size, can be reliable. Upon taking in into
account the difference in size and shape of the electrode compared to the oscillating
sphere used by Niemetz~\cite{niemetz04}, the estimated heat input due to the production
of quantum turbulence by the electrode moving at 3~cm/s is a factor of $10^7$ larger
than the sphere with the same rms velocity, but nonetheless is somewhat less
than 1~mW,
an acceptable rate for the nEDM apparatus.


%% file: discussion.tex
%
A Cavallo multiplier may be a good option for experiments that
require a very high voltage but very low current, such
as electric dipole moment experiments.
Any significant leakage current can be addressed by periodically
transferring additional charge to the load capacitor, with the
charging voltage $V_A$ adjusted to stabilize the load voltage $V_C$.
If the experiment is performed as a series of measurement cycles
that include a preparation period (e.g., loading particles into a
measurement cell) and a separate measurement period,
the charging cycle can be executed outside the measurement time interval.
The voltage during the measurement period will slowly decrease in magnitude,
depending on the leakage current, but will have zero ripple,
a possible advantage over other methods of high voltage production
in sensitive experiments.

The Cavallo multiplier is a simple device amenable
to good electrostatic design, and it can be made compatible with a cryogenic
environment.  In the case of a liquid helium filled vessel,
the device makes good use of the high dielectric strength of the medium~\cite{mshv}.
An initial design study indicates that the Cavallo multiplier is feasible
for a large-scale, cryogenic experiment to measure the neutron
electric dipole moment.
Another possible application is in
time projection chambers (TPCs)~\cite{Marx1978}
operated in liquid nobles such as argon and xenon~\cite{Rubbia2013,Aprile2012}.
An electric field of 500--1000~V/cm is needed to drift electrons
generated by passage of charged particles to the detector grid.
Currently, the length of the drift region is limited
by the performance of the high voltage feedthrough~\cite{Rebel2014,Cantini2017}.
Liquid nobles, which include helium, have high dielectric strengths,
and similar electrostatic design constraints apply as in the Cavallo multiplier
for the nEDM experiment (Sec.~\ref{sec:cavallo_in_nEDM}).

A TPC requires a constant current to flow
from the anode to the cathode through the field cage resistor chain to maintain
a uniform electric field in the drift region.
The amount of current flowing in the resistor chain must be much greater
than the expected ionization current in the drift region.
Depending on the current requirement and other specifications like voltage
stability, a Cavallo multiplier periodically executing charging cycles
to top off the load capacitance $C_{CG}$ may not be sufficient.
One could imagine arranging several Cavallo multipliers in parallel to increase
the supply current, with the charging cycles run out of phase to even out
the voltage on the shared load capacitor.
However, an arrangement similar to the high voltage generator of Pelletron
accelerators~\cite{herb74}, which is essentially a continuous version of a Cavallo multiplier
in which plate B is divided into many small segments and arranged to operate
sequentially, may more suited for this application. 

%% file: cavallo.bbl
\begin{thebibliography}{99}

\bibitem{cavallo1795} T. Cavallo, \textit{A Complete Treatise on Electricity}, Vol. III,
 4th Edition (C. Dilly, London, 1795), 98--107.

\bibitem{herb74} R.G. Herb, ``Pelletron accelerators for very high voltage,'' Nucl. Instrum. Meth. {\bf 122}, 267--276 (1974).

\bibitem{comsol} COMSOL, Inc., \textit{COMSOL Multiphysics Reference Manual, version 5.3}, \url{http://www.comsol.com}.

\bibitem{horenstein2001} M.N. Horenstein and P.R. Stone, ``A micro-aperature electrostatic
field mill based on MEMS technology,'' J. Electrostatics, {\bf 51--52}, 515--521 (2001).

\bibitem{gollam} R. Golub and S.K. Lamoreaux, ``Neutron electric-dipole moment,
ultracold neutrons and polarized $^3$He,'' Phys. Rep. {\bf 237} (1), (1994) 1--62.

\bibitem{ito_nedm} T.M. Ito, ``Plans for a neutron EDM experiment at SNS,''
 J. Phys.: Conf. Ser. {\bf 69}, 012037 (2007).

\bibitem{mshv} T.M. Ito et al., ``An apparatus for studying electrical breakdown
in liquid helium at 0.4 K and testing electrode materials for the neutron electric
dipole moment experiment at the Spallation Neutron Source,''
Rev. Sci. Instrum. {\bf 87}, 045113 (2016).

\bibitem{nonmag_ft} Off-the-shelf commercial feedthroughs may not
be suitable for the nEDM apparatus due to strict limits on magnetic materials,
but a similar, custom feedthrough using only non-magnetic materials appears feasible.

\bibitem{rogowski} W. Rogowski, ``Die elektrische Festigkeit
 am Rande des Plattenkondensators. Ein Beitrag zur Theorie der
 Funkenstrecken um Durchf{\"u}lhrungen,''
 Archiv f{\"u}r Electrotechnik {\bf 12} (1), 1-15 (1923).

\bibitem{cobine} J.D. Cobine, \textit{Gaseous Conductors: Theory and Engineering
 Applications} (Dover, 1958, reprint of McGraw-Hill, 1941), 177--181.

\bibitem{Vinen} W.F. Vinen, ``Quantum turbulence: achievements and challenges,'' J. Low Temp. Phys. {\bf 161}, 419--444 (2010).

\bibitem{niemetz04} M.~Niemetz and W.~Schoepe, ``Stability of laminar and turbulent flow
 of superfluid $^4$He at mK temperatures around an oscillating microsphere,''
 J. Low Temp. Phys. {\bf 135}, 447--469 (2004).

\bibitem{Kopnin} N.B. Kopnin, ``Vortex instability and the onset of superfluid turbulence,''
 Phys. Rev. Lett. {\bf 92}, 135301 (2004).

\bibitem{Marx1978} J.N. Marx and D.R. Nygren, ``The Time Projection Chamber,''
 Physics Today {\bf 31} (10), 46--53 (1978).

\bibitem{Rubbia2013} A. Rubbia, ``Future liquid Argon detectors,''
  Nucl. Phys. B (Proc. Suppl.) {\bf 235--236} 190--197 (2013).

\bibitem{Aprile2012} E. Aprile et al., ``The {XENON100} dark matter experiment,''
  Astroparticle Physics {\bf 35}, 573--590 (2012).

\bibitem{Rebel2014} B. Rebel et al., ``High voltage in noble liquids for high energy physics,''
  JINST {\bf 9} (08), T08004 (2014).

\bibitem{Cantini2017} C. Cantini et al., ``First test of a high voltage feedthrough for
 liquid Argon {TPC}s connected to a 300 kV power supply,'' JINST {\bf 12}, P03021 (2017).


\end{thebibliography}
